\documentclass[pra,twocolumn]{revtex4}
\usepackage{amsthm}
\usepackage{amsmath}
\usepackage{latexsym}
\usepackage{amsfonts}
\usepackage{amssymb}
\usepackage{color}
\usepackage{bbm,dsfont}
\usepackage{graphicx}
\usepackage{hyperref}
\usepackage{subfigure}
\usepackage{caption}
\usepackage{xr}

\newtheorem{proposition?}{Proposition?}

\theoremstyle{definition}

\begin{document}
\title[]{Einstein-Podolsky-Rosen steering based on semi-supervised machine learning}
\author{Lifeng Zhang, Zhihua Chen}
\thanks{Electronic address:  chenzhihua77@sina.com}
\affiliation{School of Science, Jimei University, Xiamen 361021,China}

\author{Shao-Ming Fei}
\thanks{Electronic address:  feishm@cnu.edu.cn}
\affiliation{School of Mathematical Sciences, Capital Normal University, Beijing 100048, China,\\
Max Planck Institute for Mathematics in the Sciences, 04103 Leipzig, Germany}

\begin{abstract}
Einstein-Podolsky-Rosen(EPR)steering is a kind of powerful nonlocal quantum resource in quantum information processing such as quantum cryptography and quantum communication. Many criteria have been proposed in the past few years to detect the steerability both analytically and numerically. Supervised machine learning such as support vector machines and neural networks have also been trained to detect the EPR steerability. To implement supervised machine learning, one needs a lot of labeled quantum states by using the semidefinite programming, which is very time consuming. We present a semi-supervised support vector machine method which only uses a small portion of labeled quantum states in detecting quantum steering. We show that our approach can significantly improve the accuracies by detailed examples.
\end{abstract}

\maketitle

\section{Introduction}
Einstein-Podolsky-Rosen (EPR) steering is an intermediate quantum correlation between quantum entanglement and quantum nonlocality \cite{Wiseman},
wherein one party can steer the state of another distant party by making measurements on one party of a shared bipartite state.
EPR-steering was first observed by Schr\"odinger in the famous EPR paradox in 1935 \cite{Schrodinger,Einstein}.
However, the rigorous definition of EPR-steering was not proposed till 2007. EPR-steering is proven to be asymmetric in general, that is, one party can steer the another but not vice versa \cite{Bowles, A-V-N, Bowles2}. It has been shown that steering has many applications in quantum information processing such as one-sided device independent quantum key distribution \cite{Branciard}, randomness certification \cite{Passaro,Coyle}, subchannel discrimination \cite{Piani,Sun}, secret sharing \cite{Xiang}, quantum teleportation \cite{He}, coupled qubits and magnetoreception \cite{Ku}, no-cloning of quantum steering \cite{Chiu} and quantum networks \cite{Chen}.

Many methods have been proposed to detect EPR-steering, such as linear and nonlinear steering inequalities \cite{ineq1,ineq2,ineq3,ineq4,ineq5,ineq6}, uncertainty relations \cite{unc1,unc2,unc3}, moment matrix approach \cite{mom} and all-versus-nothing method \cite{A-V-N}. Many steering measures were also proposed, which can be estimated based on semidefinite programming \cite{SDP}. Nevertheless, it is generally very time consuming to detect steerability of a state numerically, as a huge amount of measurement directions has to be run over. Efficient criteria of steerability are still far from being satisfied.

Machine learning is a useful tool for classification problems, which has already successful applications in quantum information processing tasks such as entanglement classification, nonlocality discriminant, Hamiltonian learning and phase transition identification \cite{svm1,svm2,svm3,svm4,svm5,svm6}.
Machine learning methods such as support vector machine (SVM), artificial neural networks and decision trees have been also applied to the EPR-steering detection and quantification \cite{Changliang, Zhang}. However, these methods are supervised machine learning ones, in which a lot of labeled results are needed. While quantum state labeling is also a very time consuming task.

Semi-supervised machine learning was proposed to combine labeled and unlabeled data to improve the learning behavior \cite{Zhu X}. Semi-supervised
support vector machine combines  support vector machine and semi-supervised learning, which can make use of the scarce labeled
data and sufficient unlabeled data to improve the classification performance \cite{Ding}. The safe semi-supervised SVM (S4VM) was proposed
to exploit multiple candidate low-density separators and to select the representative separators, which is never significantly inferior to the inductive SVM \cite{S4VM}.
Semi-supervised SVMs have been successfully applied to text and image classification, face recognition and many other areas \cite{ssvm1,ssvm2,ssvm3,ssvm4}.

In this paper, we apply semi-supervised SVM-S4VM to deal with  quantum steering problems. We need
only a small portion of labeled states generated randomly by the semidefinite programming, together with a large portion of unlabeled quantum states used in S4VM.
Compared with the inductive SVM, the accuracies can be significantly improved. Moreover, less time is needed to label the quantum states, while  high accuracy is still attained.

\section{Supervised Machine Learning of EPR-Steering}

A two-qubit quantum state,
$\rho_{AB}=\frac{1}{4}(I_4+\sum\limits_{i=1}^3 r_i\sigma_i\otimes I_2+\sum\limits_{i=1}^3 s_i I_2\otimes \sigma_i+\sum\limits_{i,j=1}^3 t_{ij}\sigma_i\otimes\sigma_j)$,
where $I_d$ is the $d\times d$ identity matrix and $\sigma_i$ ($i=1,2,3$) are the standard Pauli matrices,
is said to admit a local hidden state (LHS) model if the probability $P(a,b|\textbf{A},\textbf{B})$ that Alice and Bob performs the measurements $\textbf{A}$  and $\textbf{B}$ with measurement outcomes $a$ and $b$, respectively, satisfy
\begin{eqnarray}\label{1}\nonumber
P(a,b|\textbf{A},\textbf{B})=&\rm{tr}[(M_{\textbf{A}}^a\otimes M_{\textbf{B}}^b).\rho_{AB}]\\
=&\sum p(\lambda)p(a|\textbf{A},\lambda)p_Q(b|\textbf{B},\lambda)
\end{eqnarray}
where $P_Q(b|\textbf{B},\lambda)=\rm{tr}[\rho_{\lambda} M_{\textbf{B}}^b]$, $\rho_{\lambda}$ are qubit states specified by the parameter $\lambda$.
State $\rho_{AB}$ is said to be not steerable if it satisfies relation (\ref{1}). Otherwise we say that $\rho_{AB}$ is steerable from Alice to Bob.
Equivalently, one may say that Alice can not steer Bob if there exist probability $p_{\lambda}$ and a set of quantum states $\rho_{\lambda}$ such that
\begin{equation}
\rho^a_{\textbf{A}}=\rm{tr}_A((M^a_{\textbf{A}}\otimes I_2).\rho_{AB})=\sum p_{\lambda}p(a|\textbf{A},\lambda)\rho_{\lambda},
\end{equation}
where $p(a|\textbf{A},\lambda)$ is a probability distribution given by $a$ and $\lambda$.

For any $\rho_{AB}$ if Alice performs $m$ measurements $\textbf{A}=\{0,1,2,\cdots,m-1\}$
with $q$ outcomes $a\in\{1,2,\cdots,q\}$, an assemblage $\{\rho_{\textbf{A}}^a\}_{\textbf{A},a}$ of $m$ ensembles is obtained.
Denote $D(a|\textbf{A},\lambda)=\delta_{a,\lambda(\textbf{A})}$, that is, $D(a|\textbf{A},\lambda)=1$ if $\lambda(\textbf{A})=a$ and $D(a|\textbf{A},\lambda)=0$ if $\lambda(\textbf{A})\neq a$. The following SDP algorithm can be used to detect if Alice can steer Bob
for any given $\rho^a_{\textbf{A}}$ and $D(a|\textbf{A},\lambda)$ \cite{SDP},
\begin{eqnarray}
\label{sdp}
&\min\limits_{F_{a|\textbf{A}}}\rm{tr}\sum\limits_{a,\textbf{A}}F_{a|\textbf{A}}\rho^a_{\textbf{A}}\\ \nonumber
&\rm{such}\ \rm{that}.\sum\limits_{a,\textbf{A}}F_{a|\textbf{A}}D(a|\textbf{A},\lambda)\geq 0,\\ \nonumber
&\rm{tr}(\sum\limits_{a,\textbf{A},\lambda}F_{a|\textbf{A}}D(a|\textbf{A},\lambda))=1. \nonumber
\end{eqnarray}
If the objective value of (\ref{sdp}) is negative for some measurements $\textbf{A},$ $\rho_{AB}$ is
steerable from Alice to Bob. Otherwise, there exists an LHS model.

A state $\rho_{AB}$ is labeled $-1$ if the objective value is negative after using the SDP program 100 times with different values of measurements.
Otherwise, the state is labeled $+1$. In \cite{Changliang} a SVM is used for the classifier.
For $m=2,3,\cdots,8$, $5000$ samples with label $+1$ and $5000$ samples with $-1$ were obtained.
The last $1000$ positive samples and $1000$ negative samples were reserved for tests, the remaining
$4000$ positive and $4000$ negative samples were kept as training set to learn a classifier.

SVM is a supervised learning model used for classification and regression analysis.
Given $(\textbf{x}_i,y_i)$, $i=1,2,\cdots,n$,
where $n$ is the number of samples, $\textbf{x}_i$ $(i=1,2,\cdots,n)$ are the feature vectors of the samples and
$y_i$ $(i=1,2,\cdots,n)$ are the labels of the samples, $y_i=1$ or $y_i=-1$.
For linearly separable problems, SVM aims to find a hyperplane with the largest distance to the nearest feature vectors of each class.
The hyperplanes $H$ are defined as $\omega \textbf{x}_i+b\geq 1$ when $y_i=1$, and $\omega \textbf{x}_i+b\leq -1$ when $y_i=-1$.
Two marginal hyperplanes are defined by $H_1:$ $\omega \textbf{x}_i+b= 1$ and $H_2:$ $\omega \textbf{x}_i+b=-1.$
The vectors on $H_1$ and $H_2$ are the support vectors. The distance $\frac{2}{\|\omega\|}$ between $H_1$ and $H_2$ needs to be
maximized. The final separating hyperlane $H$ is selected to be the middle one between the two marginal hyperplanes.

Hence, the problems to be solved are given as follows,
\begin{eqnarray}
&\min\limits_{\omega}\frac{1}{2}\omega^T\omega\\ \nonumber
\rm{such}\ \rm{ that} &y_i(\omega\textbf{x}_i+b)\geq 1. \nonumber
\end{eqnarray}
When the errors $\{{\xi_i}\}_{i=1}^n$ in classification are allowed, the soft marginal hyperplane problem is attained,
\begin{eqnarray}
&\min\limits_{\omega,b,\xi}\frac{1}{2}\omega^T\omega+C\sum\limits_{i=1}^n \xi_{i}\\ \nonumber
\rm{such}\ \rm{that} &y_i(\omega\textbf{x}_i+b)\geq 1-\xi_{i},\\ \nonumber
&\xi_i\geq 0,
\end{eqnarray}
where $\xi_i$ are slack variables, $\xi_i=0$ if there is no error for $\textbf{x}_i$,
$C$ is a tradeoff parameter between the error and the margin.
For non-linearly separable problems, by using the kernel trick the problem needing to be solved becomes
\begin{eqnarray}
\label{svm}
&\min\limits_{\omega,b,\xi}\frac{1}{2}\omega^T\omega+C\sum\limits_{i=1}^n \xi_{i}\\ \nonumber
\rm{such}\ \rm{that} &y_i(\omega\phi(\textbf{x}_i)+b)\geq 1-\xi_{i},\\ \nonumber
&\xi_i\geq 0,
\end{eqnarray}
where
$\phi(\textbf{x})$ is the kernel function \cite{svmp1,svmp2}.

Concerning the steering detection problem, the training set $\{\textbf{x}_i,y_i\}$ $(i=1, 2, \cdots, 8000)$ and the radial basis function
$K(\phi(\textbf{x}_i),\phi(\textbf{x}_j))=e^{-\gamma\|\textbf{x}_i-\textbf{x}_j\|^2}$ are used with the parameters $C$ and
$\gamma$ determined by a grid search approach. The classifier is attained by solving (\ref{svm}) \cite{Changliang}.

\section{Semi-supervised Machine Learning of EPR-Steering}
A well-trained SVM needs a lot of labeled samples. For steering detection it takes much time to generate the labeled
samples and to obtain the data set with the increase of measurements \cite{Changliang}.
Semi-supervised SVM can be used in  classifying problems with a small portion of labeled
samples and a large number of unlabeled samples.

The S3VM was first proposed to simultaneously learn the optimal hyperplane
and the labels for unlabeled instances \cite{s3vm, Zhu}.
Given a set of $l$ labeled data $\{\textbf{x}_i, y_i\}_{i=1}^l$ and a set of $u$ unlabeled data $\{\hat{x}_j\}_{j=l+1}^{l+u}$, $y_i\in\{1,-1\}$,
one needs to solve the following optimization problem:
\begin{eqnarray}
\label{s3vm}
&\min\limits_{\omega,b,\hat{y}\in{\emph{B}}}(\frac{1}{2}\|\omega\|^2+C_1\sum\limits_{i=1}^l \xi_i+C_2\sum\limits_{j=l+1}^{l+u} \hat{\xi}_j)\\ \nonumber
& \rm{such}\ \rm{that} \quad y_i(\omega'\phi(\textbf{x}_i)+b)\geq 1-\xi_i,~~ \xi_i\geq 0\\ \nonumber
&\hat{y}_{j}(\omega'\phi(\hat{\textbf{x}}_j)+b)\geq 1-\hat{\xi}_j,~~ \hat{\xi}_j\geq 0 \\ \nonumber
&\forall i=1,\cdots,l,~\forall j=l+1,\cdots, l+u,
\end{eqnarray}
where $\emph{B}=\{\hat{y}\in \{\pm 1\}^u|-\beta\leq \frac{\sum\limits_{j=l+1}^{l+u}\hat{y}_{j}}{u}-\frac{\sum\limits_{i=1}^l y_i}{l}\leq \beta\}$
is the balanced constraint to avoid the trivial solution.

Unlike S3VM, the S4VM was proposed to construct diverse large-margin separators since multiple large-margin low density separators may coincide with
the limited labeled data, and then the label assignment for unlabeled instances is optimized such that the worst-case performance improvement
over inductive SVM is maximized \cite{S4VM}. Recalling the S4VM in detail, one first has the following minimization problem:


\begin{eqnarray}
\label{s4}
&\min\limits_{\{\omega_t,b_t,\hat{y}_t\in{\emph{B}}\}_{t=1}^T}\sum\limits_{t=1}^T(\frac{1}{2}\|\omega_t\|^2+C_1\sum\limits_{i=1}^l \xi_i+C_2\sum\limits_{j=l+1}^{l+u} \hat{\xi}_j)\\ \nonumber
&+G \sum\limits_{1\leq t\neq \tilde{t}\leq T}\delta(\frac{\hat{y}'_t\hat{y}_{\tilde{t}}}{u}\geq 1-\varsigma),\\ \nonumber
& \rm{such}\ \rm{that}\ y_i(\omega_t'\phi(\textbf{x}_i)+b_t)\geq 1-\xi_i,~~ \xi_i\geq 0\\ \nonumber
&\hat{y}_{t,j}(\omega_t'\phi(\hat{\textbf{x}}_j)+b_t)\geq 1-\hat{\xi}_j,~~ \hat{\xi}_j\geq 0 \\ \nonumber
&\forall i=1,\cdots,l,~\forall j=l+1,\cdots, l+u,~\forall t=1,\cdots, T.
\end{eqnarray}
where $\emph{B}=\{\hat{y}_{t}\in \{\pm 1\}^u|-\beta\leq \frac{\sum\limits_{j=l+1}^{l+u}\hat{y}_{t,j}}{u}-\frac{\sum\limits_{i=1}^l y_i}{l}\leq \beta\},$ $\hat{y}_{t,j}$ is the
$jth$ entry of $\hat{y}_t,$ $\delta$ is the indicator function which represents  a quantity of
penalty about the diversity of separators and $\varsigma\in [0,1]$ is a constant, $G$ is a large constant enforcing large diversity.
$C_1$ and $C_2$
are regularization parameters trading off the complexity
and the empirical error on label and unlabeled data. $T$ is the number of the separators.

Second, from the global simulated annealing search with a deterministic local search scheme method or the sampling strategy used in solving minimization problems Eq.(\ref{s4}),
we have the following problem,
\begin{eqnarray}
\label{sl}
\bar{y}=arg \max\limits_{y\in\{\pm 1\}^u}\min\limits_{\hat{y}\in M_0} J(y,\hat{y},y^{\rm{SVM}}),
\end{eqnarray}
where $y^{\rm{SVM}}$ is the predictive labels of inductive SVM on unlabeled instances,
$M_0=\{\hat{y}_t\}_{t=1}^T$ is obtained from \eqref{s4},
\begin{eqnarray}
J(y,\hat{y},y^{\rm{SVM}})=&\rm{gain}(y,\hat{y},y^{\rm{SVM}})-\lambda \rm{loss}(y,\hat{y},y^{\rm{SVM}}) \nonumber\\
=&c_t' y + d_t
\end{eqnarray}
with
\begin{eqnarray}
&\rm{gain}(y,\hat{y},y^{\rm{SVM}})=\sum\limits_{j=l+1}^{l+u}\frac{1+y_j\hat{y}_j}{2}\frac{1-y^{\rm{SVM}}_j\hat{y}_j}{2},\\ \nonumber
&\rm{loss}(y,\hat{y},y^{\rm{SVM}})=\sum\limits_{j=l+1}^{l+u}\frac{1-y_j\hat{y}_j}{2}\frac{1+y^{\rm{SVM}}_j\hat{y}_j}{2}, \\ \nonumber
&c_t=\frac{1}{4}[(1+\lambda)\hat{y}_t+(\lambda-1)y^{\rm{SVM}}],\\ \nonumber
&d_t=\frac{1}{4}[-(1+\lambda)\hat{y}'_ty^{\rm{SVM}}+(1-\lambda)u].\nonumber
\end{eqnarray}
Then the minimization problem in \eqref{sl} can be reformulated as the following maximization problem,
\begin{eqnarray}
\label{st}
&\max\limits_{y,\tau} \tau \hspace{4.5cm}\\ \nonumber
\rm{such}\ \rm{that} &\tau \leq c_t' y + d_t,~ \forall t=1,2,\cdots, T;~ y\in\{\pm 1\}^u.
\end{eqnarray}

Convex linear programming problems can be solved by relaxing $y\in\{\pm 1\}^u$ to $y\in[-1,1]^u,$
and then project it back to integer solutions with minimum distance. If the function
value of the resulting integer solution is smaller than that
of $y^{\rm{SVM}}$, $y^{\rm{SVM}}$ is output as the final solution instead.



For any two-qubit quantum state $\rho_{AB},$ set $\rho_0=({I}_2\otimes \rho_B^{\frac{1}{2}})\rho_{AB}({I}_2\otimes \rho_B^{\frac{1}{2}}),$
where $\rho_B=\rm{tr}_A\rho_{AB}$. The coefficients $\tau_{kl}=\rm{tr}(\rho_0(\sigma_k\otimes\sigma_l))$
can be reformulate as a nine-dimensional feature vector $\textbf{x}=[\tau_{11},\tau_{12},\cdots,\tau_{33}]^T$, where
$T$ denotes transpose. Generating randomly quantum states by SDP \cite{SDP, Changliang}, we get a set of $l$ feature vectors of labeled quantum states, $\{\textbf{x}_i\}$, $i=1,2,\cdots,l$, with $y_i=-1$ for steerable states and $y_i=1$ for unsteerable states. Denote $\{\hat{\textbf{x}}_j\}_{j=l+1}^{l+u}$ the set of
feature vectors of unlabeled quantum states whose labels can be determined by  the above S4VM \cite{S4VM} method.
Radial basis function kernels are used here and $K(\phi(\textbf{x}_i),\phi(\textbf{x}_j))=e^{-\gamma\|\textbf{x}_i-\textbf{x}_j\|^2}$.
We have three hyperparameters $C_1$, $C_2$, and $\gamma$ which are also determined by the grid research method.

\section{Numerical Results}

We generate randomly quantum states and get the class of steerable states from such random states by SDP.
For balance, we choose half positive quantum states and half negative random states.

We implement SVMs by using $l$ labeled quantum states, the radial basis function kernel,
ten-folded cross validation, and the grid search approach. $u$ unlabeled quantum states are considered as the test sets and the test errors are obtained.
For S4VM, sample strategy is implemented here, three hyperparameters $C_1$, $C_2$, and $\gamma$ are determined by the grid search approach. We set $\beta=0.1, \lambda=3,$ the number
 of clusters $T$ is 10 and the number of samples is 100 in the sample strategy for the examples.
$u$ unlabeled quantum states are divided into $M$ different sets. Taking $M=2$ as an example, we have the following steps:

$(1)$ The first set of unlabeled quantum states can be predicted by using $l$ labeled quantum states.
Then, we implement five-fold cross validation for these $l+u/M$ quantum states by using SVMs. The
best cross-validation accuracy and the best hyperparameters are then obtained. The labels of the first set of unlabeled quantum states from the best hyperparameters
can be regarded as the real labels and the classification accuracies of these unlabeled quantum states are obtained.

$(2)$ After labeling the first set of unlabeled quantum states, $l+u/M$ labeled quantum states and the second set of unlabeled quantum states can be utilized to
implement S4VM the second time. Then the second set of unlabeled quantum states can also be labeled.
Implementing five-fold cross validation for these $l+2*u/M,$ quantum states by using SVMs,
the best cross-validation accuracy and the best hyperparameters are obtained.
The labels of the second unlabeled quantum states from the best hyperparameters are considered as the real labels
and the classification accuracies of the second set of the unlabeled quantum states are obtained.

$(3)$ The average accuracy of the two sets of unlabeled quantum states is considered as the classification accuracy of the unlabeled set.

\begin{figure}
\centering
 \includegraphics[width=6cm]{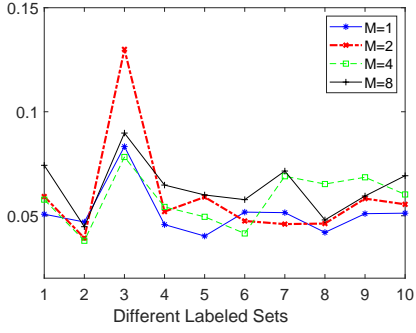}
 \label{fig:subfig:a}
\caption{Classification errors of the unlabeled quantum states by using ten different sets of $l$ labeled quantum states for $m=2,$ $l=30,$ and
$u=4000$. The errors of the unlabeled states by S4VM for $M=1, 2, 4, 8$ are represented by solid line with star, thick dashed line with x, dashed line with square and solid line with $+$, respectively.
}
\label{fig1}
\end{figure}

\begin{figure}
\centering
 \includegraphics[width=6cm]{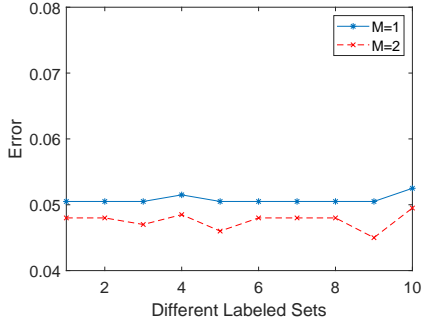}
 \label{fig:subfig:a}
\caption{Classification errors of the unlabeled quantum states by using ten different sets of 30 labeled quantum states for $m=8$ and
$u=4000$. The errors of the unlabeled states by S4VM for $M=1$ and $2$ are represented by solid line with star and thick dashed line with x, respectively.
}
\label{fig2}
\end{figure}

We compute the errors of $4000$ unlabeled quantum states when $30$ labeled quantum states are used to implement S4VM for $m=2$
and $M= 1, 2, 4, 8$. Except for the third labeled set, the errors are smaller when $M=1$ and $M=2$ compared to the errors when $M=4$ and $M=8$ for the most cases
but the errors when $M=2$ are smaller than that when $M=1$ for $m=8$, see Figs. \ref{fig1} and \ref{fig2}.
In addition, the computational speed for $M=2$ is three times faster than that for $M=1.$

\begin{figure}
\centering
\includegraphics[width=8cm]{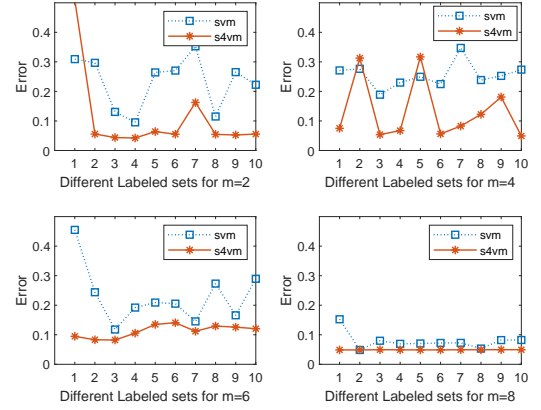}
\caption{Classification errors of SVM and S4VM by using ten different sets of $l$ labeled quantum states for $m=2, 4,6,$ and $8,$
$l=10,$ $u=4000$ and $M=2$. The accuracies of the unlabeled states by SVM and S4VM are represented by the lines with squares and $\star$, respectively.}
\label{fig3}
\end{figure}

\begin{figure}
\centering
\includegraphics[width=8cm]{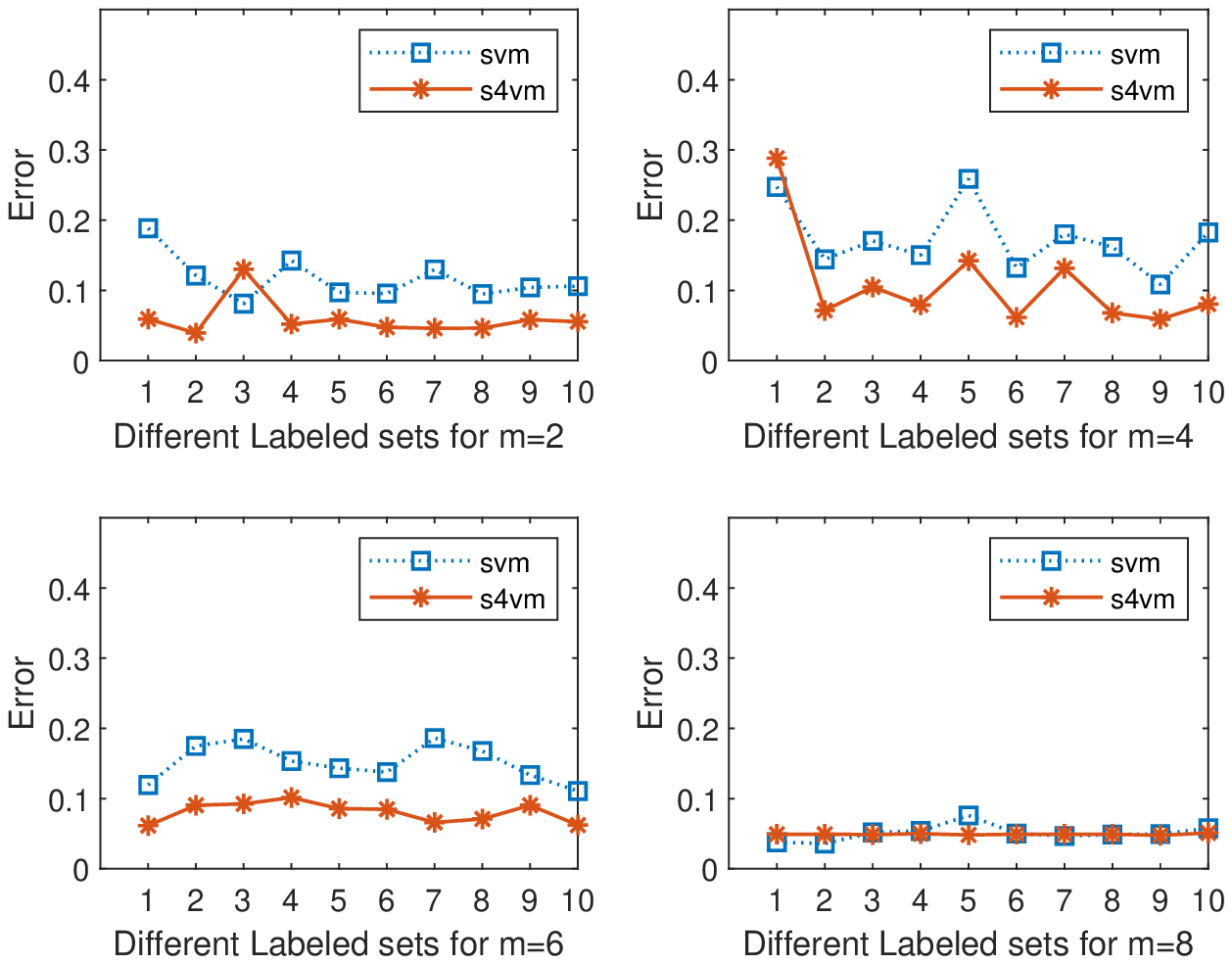}
\caption{Classification errors of SVM and S4VM by using ten different sets of $l$ labeled quantum states for $m=2, 4,6,$ and $8,$ $l=30,$
$u=4000$ and $M=2$. The accuracies of the unlabeled states by SVM and S4VM are represented by the lines with squares and $\star$, respectively,
}
\label{fig4}
\end{figure}

\begin{figure}
\centering
\includegraphics[width=8cm]{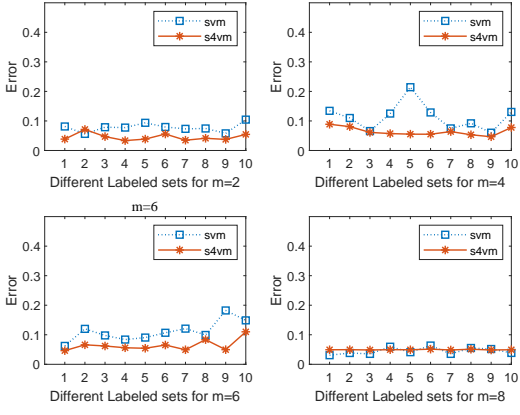}
\caption{Classification errors of SVM and S4VM by using ten different sets of $l$ labeled quantum states for $m=2, 4,6,$ and $8,$ $l=50,$
$u=4000$ and $M=2$. The accuracies of the unlabeled states by SVM and S4VM are represented by lines with squares and $\star$, respectively.}
\label{fig5}
\end{figure}

Based on the above errors and computational speed, in the following numerical experiments
we set $M=2$ and use $10$, $30$, or $50$ labeled states and $4000$ unlabeled quantum states to implement S4VM for $m=2, 4, 6, 8$.
When Alice performs $m$ $(m=2,4,6,8)$ measurements, S4VM is implemented ten times by using ten different sets of $l$ labeled quantum states.
The errors for these $u$ unlabeled quantum states are shown in the Figs. \ref{fig3}-\ref{fig5}.

\begin{figure}
\centering
\includegraphics[width=8cm]{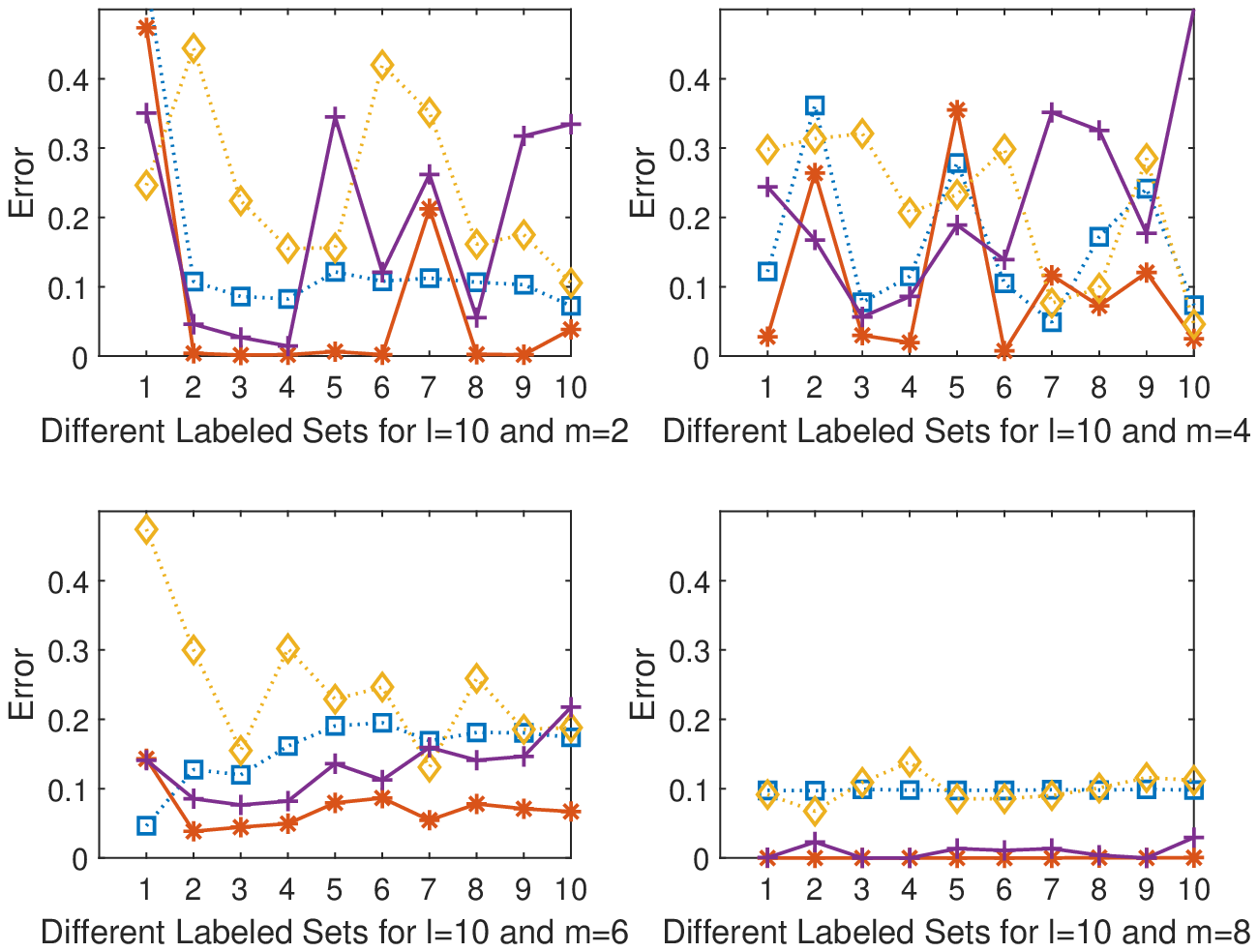}
\caption{Classification errors of $2000$ positive quantum states and $2000$ negative quantum states by
SVM and S4VM with ten different labeled sets for $l=10,$  $m=2, 4,6,$ and $8$.
The errors for positive and negative quantum states by S4VM are represented by lines with squares and $\star$, respectively,
while the errors for positive and negative quantum states by SVM are represented by lines with diamond and $+$, respectively.
}
\label{fig6}
\end{figure}

\begin{figure}
\centering
\includegraphics[width=8cm]{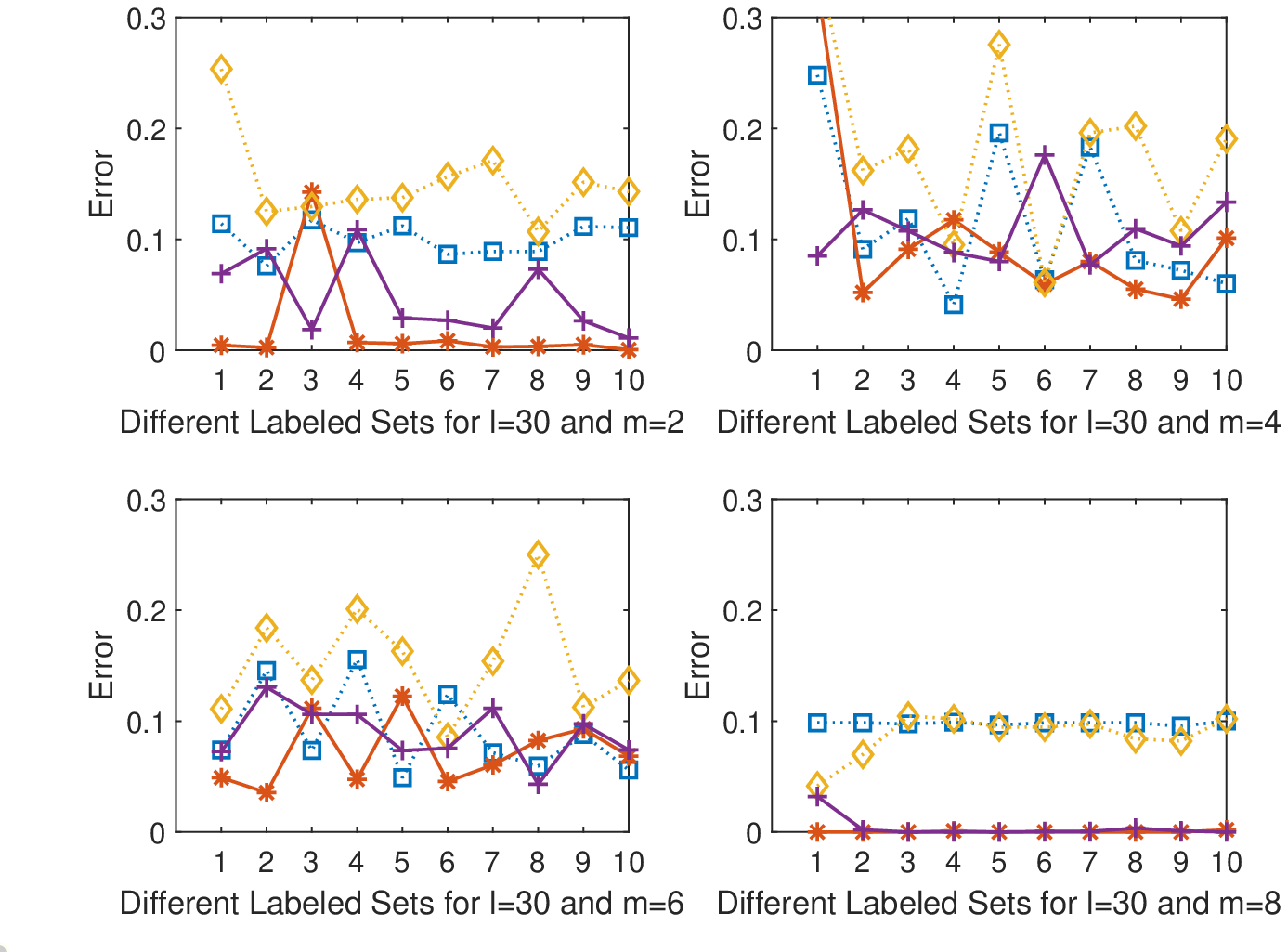}
\caption{Classification errors of $2000$ positive quantum states and $2000$ negative quantum states by
SVM and S4VM with ten different labeled sets for $l=30,$ $m=2, 4,6,$ and $8$.
The errors for positive and negative quantum states are represented by lines with squares and $\star$,
respectively, while the errors for positive and negative quantum states by SVM are represented by lines with diamond and $+$, respectively.
}
\label{fig7}
\end{figure}

\begin{figure}
\centering
\includegraphics[width=8cm]{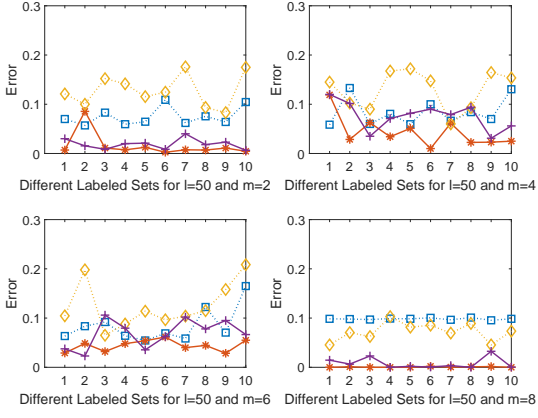}
\caption{Classification errors of $2000$ positive quantum states and $2000$ negative quantum states by
SVM and S4VM with ten different labeled sets for $l=50,$ $m=2, 4,6,$ and $8$.
The errors for positive and negative quantum states are represented by lines with squares and $\star$,
respectively, while the errors for positive and negative quantum states by SVM are represented by lines with diamond and $+$, respectively.
}
\label{fig8}
\end{figure}

When two measurements are performed and ten labeled quantum states are used, the errors by SVM are larger than $0.2$ in many cases, which can
be reduced to about 0.05 by S4VM in most cases.
When $l=30$, the errors by S4VM are less than 0.06, while the errors by SVM are larger than 0.09 in most cases. When $l=50,$
the errors by S4VM are less than 0.05, while the errors by SVM are larger than 0.07 in most cases.
In Table \ref{tab1}, we list the maximum differences between errors from SVM and S4VM  by
using 10, 30, and 50 labeled quantum states and 4000 unlabeled quantum states
for $m=2,\,4,\,6$, and 8. It can be seen that the accuracies are significantly improved.
\begin{table}
\caption{The maximum differences between the errors from SVM and S4VM by using ten different $l$ labeled quantum states and 4000 unlabeled quantum states for $l=10,30,$ and $50$ and $m=2,4,6,$ and $8$. $\Delta E_{\rm{max,l}}$ is the maximum difference, the errors from SVM minus the errors from S4VM, by using $l$ labeled quantum states.}
\label{tab1}
\begin{tabular}{|l|l|l|l|}
\hline
$m$ &$\Delta E_{\rm{max,10}}$&$\Delta E_{\rm{max,30}}$&$\Delta E_{\rm{max,50}}$\\ \hline
2   & 0.216& 0.129 & 0.055 \\ \hline
4   & 0.22 & 0.088 & 0.159 \\ \hline
6   & 0.36 & 0.12  & 0.13 \\ \hline
8   & 0.104& 0.028 & 0.01\\ \hline
\end{tabular}
\end{table}
To investigate the performance on  positive and negative quantum states by S4VM, we show the errors of positive class and negative class for $l=10$, $30$ and $50$ in FIGs. \ref{fig6}, \ref{fig7} and \ref{fig8}.

The errors for positive and negative classes are also reduced in most cases, i.e. the errors for either the positive or negative class by SVM are also larger than the corresponding ones by S4VM for $l=10$, $30$, and $50$, respectively.
In general the errors for the negative class are smaller than those for the positive class, maybe due to the errors of SDP.
When $m=2$ and $m=8$ the errors by S4VM for the negative class are approximately zero in most cases.

In Table \ref{tab2} (Table \ref{tab3}), we also list the maximum differences between errors of positive
(negative) states from S4VM and SVM by using 10, 30, or 50 labeled quantum states
and 4000 unlabeled quantum states for $m=2,4,6,$ and 8.
\begin{table}
\caption{The maximum differences between the errors of positive states from SVM and S4VM by using ten different $l$ labeled quantum states and 4000 unlabeled quantum states for $l=10,30,$ and $50$ and $m=2,4,6,$ and $8$. $\Delta E^{+}_{\rm{max,l}}$ is the maximum difference, errors of positive states from SVM minus the errors from S4VM, by using $l$ labeled quantum states.}
\label{tab2}
\begin{tabular}{|l|l|l|l|l|l|l|}
\hline
$m$ &$\Delta E^{+}_{\rm{max,10}}$&$\Delta E^{+}_{\rm{max,30}}$&$\Delta E^{+}_{\rm{max,50}}$\\ \hline
2   & 0.337& 0.14 & 0.11  \\ \hline
4   & 0.24 & 0.13 & 0.11 \\ \hline
6   & 0.43 & 0.19  & 0.11 \\ \hline
8   & 0.04 & 0.007 & 0.004 \\ \hline
\end{tabular}
\end{table}
\begin{table}
\caption{The maximum differences between the errors of negative states from SVM and S4VM by using ten different $l$ labeled quantum states and 4000 unlabeled quantum states for $l=10,30,$ and $50$ and $m=2,4,6,$ and $8$. $\Delta E^{-}_{\rm{max,l}}$ is the maximum difference, the errors of negative states from SVM minus the errors from S4VM, by using $l$ labeled quantum states.}
\label{tab3}
\begin{tabular}{|l|l|l|l|l|l|l|}
\hline
$m$ &$\Delta E^{-}_{\rm{max,10}}$&$\Delta E^{-}_{\rm{max,30}}$&$\Delta E^{-}_{\rm{max,50}}$\\ \hline
2   & 0.34& 0.10 & 0.03  \\ \hline
4   & 0.47 & 0.12 & 0.07 \\ \hline
6   & 0.15 & 0.10  & 0.07 \\ \hline
8   & 0.029& 0.032 & 0.031 \\ \hline
\end{tabular}
\end{table}

\begin{figure}
\centering
\includegraphics[width=8cm]{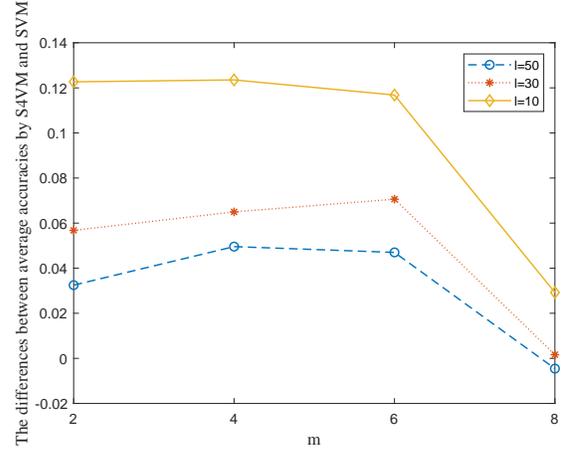}
\caption{The differences between the average errors from S4VM and SVM by using ten different $l$ labeled quantum states
and $u$ unlabeled quantum states for $m=2,4,6,$ and $8$. The differences between the average errors for $l=50,30,$ and $10$ are represented
by the lines with $\diamond$, $\star$ and circles, respectively.}
\label{fig9}
\end{figure}

Compared with the results from SVMs, semi-supervised SVMs can improve the
accuracies in most cases. The differences between average errors from S4VM and SVM by using ten different sets of labeled quantum states
are shown in Fig. \ref{fig9}. All the average errors can be improved. The smaller the $l$ (the number of labeled samples), the better the improved accuracies.

\begin{figure}
\centering
\includegraphics[width=8cm]{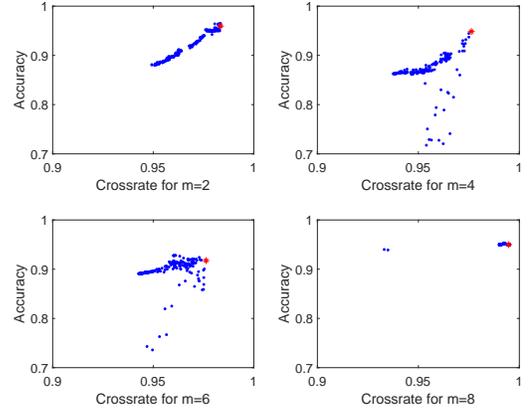}
\caption{The relationship between the cross validation accuracies and the accuracies of unlabeled quantum states by using ten labeled quantum states
and $u$ unlabeled quantum states for $m=2,4,6,$ and $8$.}
\label{fig10}
\end{figure}

To show the validity of cross validation, the relationship between the cross rate and the accuracies are shown in FIG. \ref{fig10} by taking 10
labeled quantum states. The vertical coordinates represent the accuracies of the unlabeled states, while the horizontal coordinates represent
the the accuracies of cross validation.

\begin{figure}
\centering
\subfigure[m=8]{
 \label{fig:subfig:a}
  \includegraphics[width=4cm]{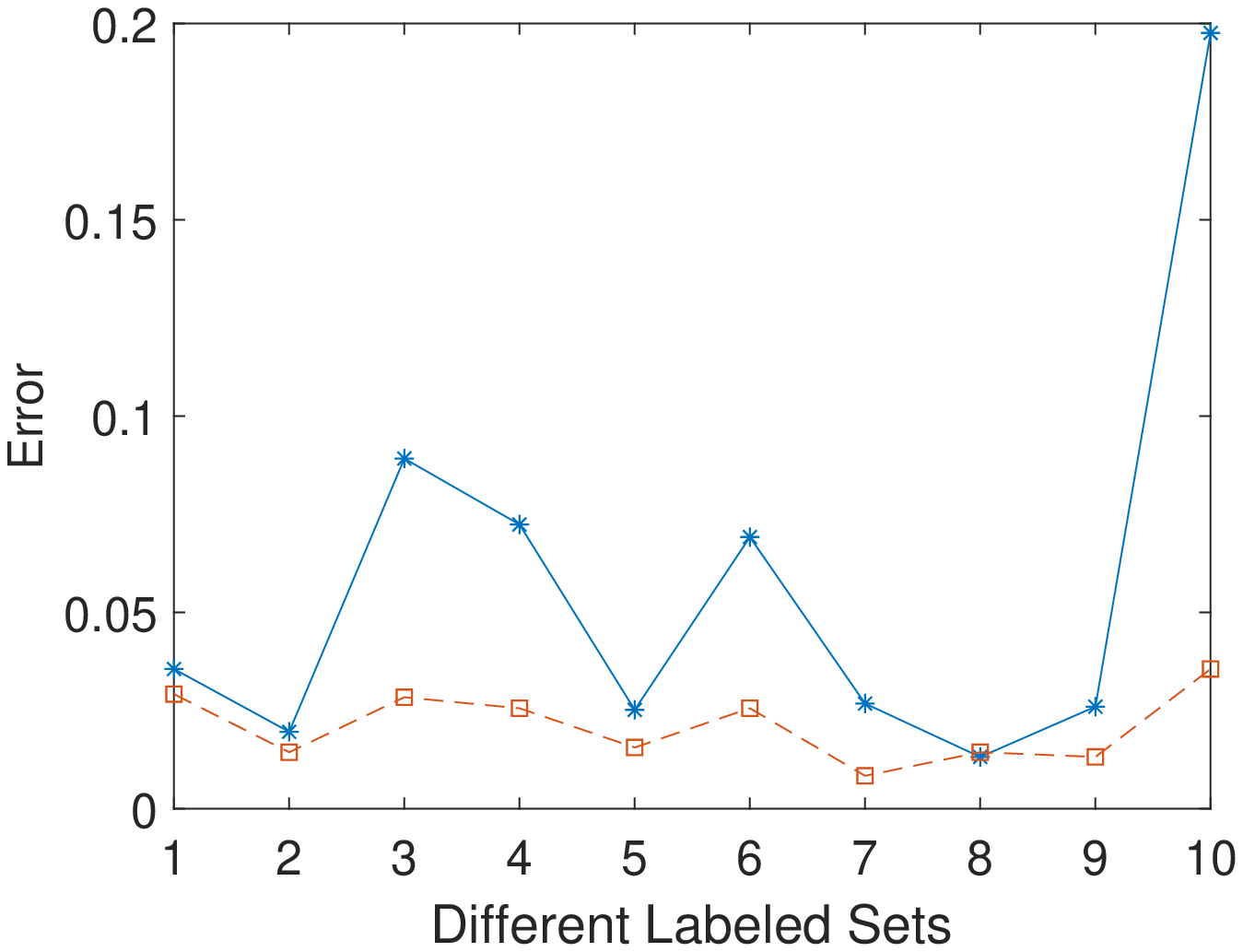}
  }
\subfigure[m=4]{
 \label{fig:subfig:b}
 \includegraphics[width=4cm]{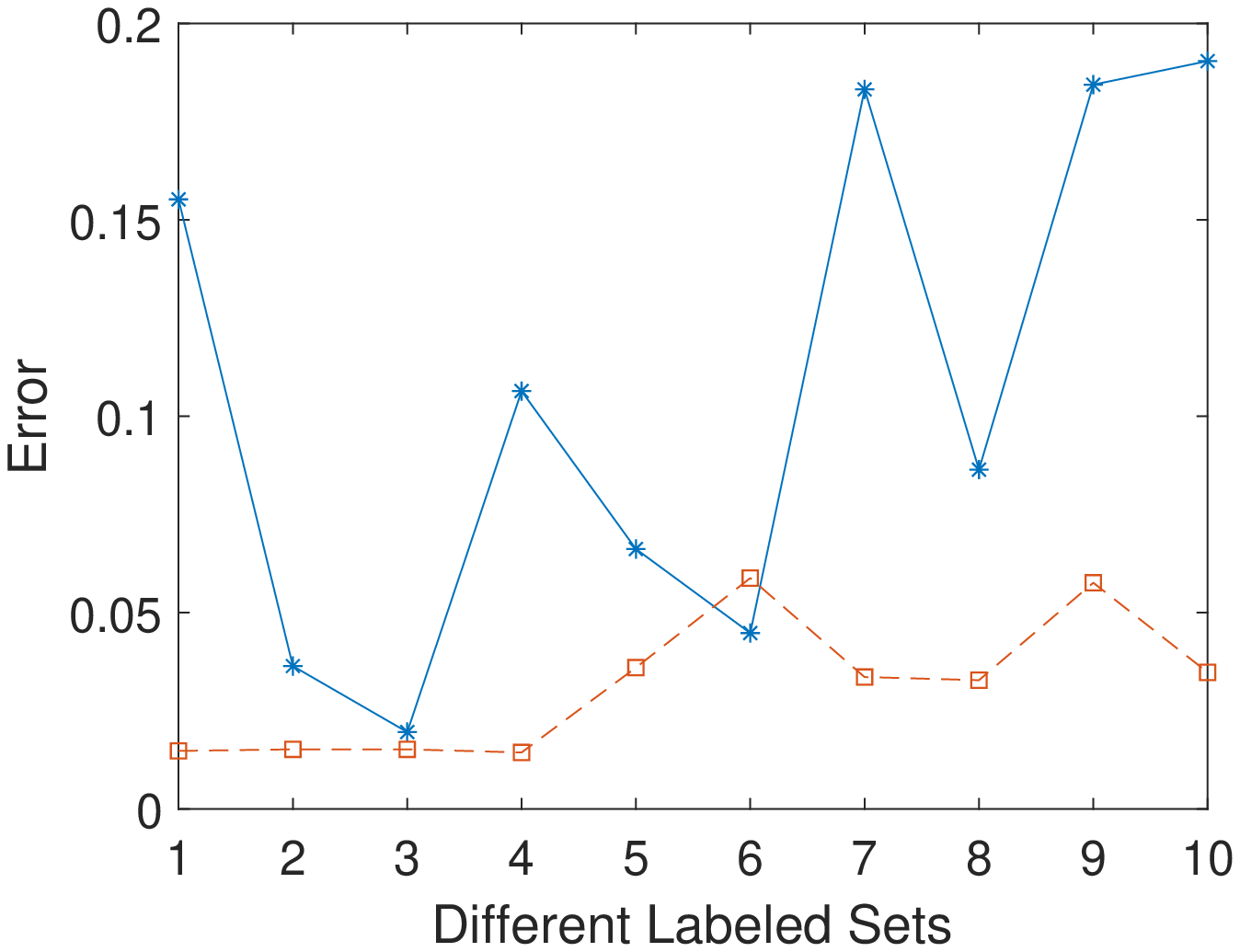}
 }
\caption{Classification errors of SVM and S4VM with ten different sets of $l$ labeled quantum states for $m=4,$ and $m=8,$ $l=30$,
$N_2=2500,$ and $M=1$. The errors of SVM for $l=30$ are represented by line with $\star$,
the errors of S4VM for $l=30$ are represented by line with square.}
\label{fig11}
\end{figure}

To investigate the validity of the S4VM, we also study the classification of the generalized Werner states,
\begin{equation}
\rho_w=p|\psi\rangle\langle\psi|+(1-p)\rho_A\otimes \frac{\rm{I}_2}{2},
\end{equation}
where $|\psi\rangle=\cos\xi|00\rangle+\sin\xi|11\rangle,$ $\rho_A=\rm{tr}_B(|\psi\rangle\langle\psi|).$
The state is unsteerable from Alice to Bob if
\begin{equation}
\cos^22\xi\geq\frac{2p-1}{(2-p)p^3}.
\end{equation}
Let the unlabeled set be of $2500$ generalized Werner states from $p=0$ to $p=1.$
By using 30 random labeled  quantum states, the classification errors of the unlabeled
generalized Werner states for $\psi=\frac{\pi}{8}$ by SVM and S4VM are shown in Fig. \ref{fig11}.
Different from the above examples, we take $M=1$ here to show the performance of S4VM.
It can be seen that the accuracies of the unlabeled quantum states can be improved to 0.95 or higher.
The accuracies by S4VM are 0.15  higher than those by SVM in the best case when $m=4$ and $m=8$.
The average accuracies by S4VM are 0.969 and 0.979 for $m=4$ and $m=8$, respectively, which are again 0.076 and 0.036 higher than those by SVM.

We only need a small portion of the labeled data to implement the semi-supervised learning algorithm, which can indeed save a lot of time.
However, the algorithm S4VM may still cost much time since the global simulated annealing search, the sample strategy, and the grid search approach in S4VM cost time. Besides, the accuracies by S4VM cannot be improved significantly when the number of the labeled samples becomes larger, while the accuracies by SVM are greater than 0.95.

\section{Conclusion}

Semi-supervised SVMs can be used in the situation that the labeled samples are scarce or
very difficult to obtain. We have implemented a special semi-supervised SVM-S4VM to
detect the EPR steering with a small portion of labeled quantum states and a large portion of unlabeled quantum states.
Compared with inductive SVM, the detection errors can be significantly decreased in most cases. Our approach makes
a useful step to solve the quantum correlation detection problems based on the semi-supervised machine learning method, as
it costs time to get enough labeled quantum states. This approach may be applied similarly to detect other quantum
correlations such as quantum entanglement and bell non-locality. Our results may also highlight the applications of
other more efficient semi-supervised machine-learning methods such as artificial neural networks to quantum correlation detections.

\bigskip

\noindent{\bf ACKNOWLEDGEMENTS}\, \, This work is supported by the National Natural Science Foundation of China (NSFC) under Grant Nos. 11571313, 12071179, 12075159 and 12171044;
Beijing Natural Science Foundation (Grant No. Z190005); Academy for Multidisciplinary Studies, Capital Normal University; the Academician Innovation Platform of Hainan Province; and Shenzhen Institute for Quantum Science and Engineering, Southern University of Science and Technology (No. SIQSE202001).


\begin{thebibliography}{99}
\bibitem{Wiseman} H. M. Wiseman, S.J. Jones , A. C. Doherty,  Steering, entanglement, nonlocality, and the Einstein-Podolsky-Rosen paradox. \emph{Phys. Rev. Lett.}, 98:140402(2007).
\bibitem{Schrodinger} E. Schrodinger, Discussion of probability relations between separated systems. \emph{Math. Proc. Camb. Philos. Soc}. 31:555(1935).
\bibitem{Einstein} A. Einstein, B. Podolsky,  N. Rosen, Can quantum-mechanical description of physical reality be considered complete? \emph{Phys. Rev.} 47:777(1935).
\bibitem{Bowles} J. Bowles, T. Vertesi, M. T. Quintino, N. Brunner: One-way Einstein-Podolsky-Rosen steering. \emph{Phys.
Rev. Lett.} 112, 200402 (2014).
\bibitem{A-V-N} J. L. Chen, X. J. Ye, C. F. Wu, H. Y. Su, A. Cabello, L. C. Kwek, C. H. Oh, All-versus-nothing proof of Einstein-Podolsky-Rosen steering, \emph{Sci. Rep.}3,2143(2013).
\bibitem{Bowles2} J. Bowles, F. Hirsch, M. T. Quintino, N. Brunner: Sufficient criterion for guaranteeing that a two-qubit
state is unsteerable. \emph{Phys. Rev. A} 93, 022121 (2016).
\bibitem{Branciard} C. Branciard, E. G. Cavalcanti, S. P. Walborn, V. Scarani, and H. M. Wiseman, One-sided device-independent quantum key distribution: Security, feasibility, and the connection with steering, \emph{Phys. Rev. A} 85, 010301(R)(2012).
\bibitem{Passaro}E. Passaro, D. Cavalcanti, P. Skrzypczyk, A. Ac$\acute{i}$n, Optimal randomness certification in the quantum steering and prepare-and-measure scenarios, \emph{New J. Phys.} 17,113010(2015).
\bibitem{Coyle}B. Coyle, M. J. Hoban, E. Kashefi, One-sided device-independent certification of unbounded random numbers, \emph{EPTCS} 273, 14-26(2018)
\bibitem{Piani} M. Piani, J. Watrous, Necessary and sufficient quantum information characterization of
Einstein-Podolsky-Rosen steering, \emph{Phys. Rev. Lett.} 114, 060404(2015).
\bibitem{Sun} K. Sun, X.-J. Ye, Y. Xiao, X.-Y. Xu, Y.-C. Wu, J.-S. Xu, J.-L. Chen, C.-F. Li, and G.-C. Guo, Demonstration of Einstein-Podolsky-Rosen steering with enhanced subchannel discrimination, \emph{npj Quantum Inf.} 4, 12(2018).
\bibitem{Xiang} Y. Xiang, I. Kogias, G. Adesso, and Q. He, Multipartite gaussian steering: monogamy constraints and quantum cryptography applications,\emph{Phys. Rev. A} 95,
010101(R)(2017).
\bibitem{He} Q. He, L. Rosales-Z$\acute{a}$rate, G. Adesso, M. D. Reid, Secure continuous variable teleportation and Einstein-Podolsky-Rosen steering,\emph{Phys.
Rev. Lett.} 115, 180502(2015).
\bibitem{Ku} H. Y. Ku, S. L. Chen, H.B Chen, N. Lambert, Y. N. Chen, F. Nori, Temporal steering in four dimensions
with applications to coupled qubits and magnetoreception. \emph{Phys. Rev. A} 94, 062126(2016).
\bibitem{Chiu} C. Y. Chiu, N. Lambert, T.L. Liao, F. Nori, C. M. Li, No-cloning of quantum steering. \emph{NPJ Quantum Inf.},
2, 16020(2016).
\bibitem{Chen} S. L. Chen, N. Lambert, C. M. Li, G. Y. Chen, Y. N. Chen, A. Miranowicz, F. Nori, Spatio-temporal steering
for testing nonclassical correlations in quantum networks. \emph{Sci. Rep.}, 7, 3728(2017).
\bibitem{ineq1} E. G. Cavalcanti, S. J. Jones, H. M. Wiseman, M. D. Reid, Experimental criteria for steering and the
Einstein-Podolsky-Rosen paradox. \emph{Phys. Rev. A} 2009, 80, 032112.
\bibitem{ineq2} M. F. Pusey, Negativity and steering: A stronger Peres conjecture. \emph{Phys. Rev. A} 2013, 88, 032313.
\bibitem{ineq3} C .L. Ren, H. Y. Su, H. F. Shi, J. L. Chen, Maximally steerable mixed state based on the linear steering
inequality and the Clauser-Horne-Shimony-Holt-like steering inequality. \emph{Phys. Rev. A} 2018, 97, 032119.
\bibitem{ineq4} Y. N. Chen, C. M. Li, N. Lambert, S. L. Chen, Y. Ota, G. Y. Chen, F. Nori, Temporal steering inequality.
\emph{Phys. Rev. A} 2014, 89, 032112.
\bibitem{ineq5} M. Zukowski, A. Dutta, Z. Yin, Geometric Bell-like inequalities for steering. \emph{Phys. Rev. A} 2015, 91, 032107.
\bibitem{ineq6} H. Zhu, M. Hayashi, L. Chen, Universal steering inequalities. \emph{Phys. Rev. Lett.} 2016, 116, 070403.
\bibitem{unc1} S. P. Walborn, A. Salles, R. M. Gomes, F. Toscano, P. H. Souto Ribeiro,  Revealing hidden Einstein-Podolsky-Rosen
nonlocality. \emph{Phys. Rev. Lett.} 2011, 106, 130402.
\bibitem{unc2} J. Schneeloch, C. J. Broadbent, S. P. Walborn,  E. G. Cavalcanti,  J.C. Howell, Einstein-Podolsky-Rosen steering
inequalities from entropic uncertainty relations. \emph{Phys. Rev. A} 2013, 87, 062103.
\bibitem{unc3} T. Pramanik, M. Kaplan, A. S. Majumdar, Fine-grained Einstein-Podolsky-Rosen steering inequalities.
\emph{Phys. Rev. A} 2014, 90, 050305.
\bibitem{mom} I. Kogias, P. Skrzypczyk, D. Cavalcanti, A. Acin, and G. Adesso, Hierarchy of Steering Criteria Based on Moments for All Bipartite Quantum Systems
2015, \emph{Phys. Rev. Lett.} 115, 210401.
\bibitem{SDP} D. Cavalcanti, P. Skrzypczyk, Quantum steering: a review with focus on semidefinite programming, \emph{Rep. Prog. Phys.} 80 024001 (2017).
\bibitem{svm1} Y. C. Ma, M. H. Yung, Transforming bell's inequalities into state classifiers with machine learning, \emph{npj, Quantum. Info.,} 4,34(2018).
\bibitem{svm2} S. R. Liu, S. L. Huang, K. R. Li, et al, A separability-entanglement classifier via machine learning, \emph{Phys. Rev. A} 98, 012315 (2018).
\bibitem{svm3} D. L. Deng, X. Li and S. D. Sarma, Quantum entanglement in neural network states, \emph{Phys. Rev. X} 7, 021021 (2017).
\bibitem{svm4} M. Yang, C. L. Ren, Y. C. Ma, Y. Xiao, X. J. Ye, L. L. Song, J. S. Xu, M. H. Yung, C. F. Li, G. C. Guo, Experimental Simultaneous Learning of Multiple Non-Classical Correlations,\emph{Phys. Rev. Lett.} 123, 190401 (2019).
\bibitem{svm5} N. Wiebe, C. Granade, C. Ferrie, and D. G. Cory, Hamiltonian learning and certification using quantum resources, \emph{Phys. Rev. Lett.} 112, 190501 (2014).
\bibitem{svm6} S. S. Schoenholz, E. D. Cubuk, D. M. Sussman, E. Kaxiras, and A. J. Liu, A structural approach to relaxation in glassy liquids, \emph{Nat. Phys.} 12, 469 (2016).
\bibitem{Changliang} C. L. Ren,C. B. Chen, Steerability detection of arbitrary 2-qubit state via machine learning, \emph{Phys. Rev. A} 100, 022314 (2019).

\bibitem{Zhang} Y.Q.Zhang, L.J. Yang, Q.L. He, L. Chen, Machine learning on quantifying quantum steerability, \emph{Quan. inf. Process.}, 19(8):263(2020)
\bibitem{Zhu X} X. Zhu,  AB. Goldberg, Introduction to semi-supervised learning. \emph{Synth. Lect. Artif. Intell. Mach. Learn.}, 3(1):1-130(2009).
\bibitem{Ding} S.F. Ding, Z.B. Zhu, X. K. Zhang, An overview on semi-supervised support vector machine, \emph{Neural. Comput.  Applic,} 28(5):969-978(2017).
\bibitem{S4VM} Y.F. Li, Z.H. Zhou, Towards making unlabeled data never hurt, \emph{IEEE Transactions on Pattern Analysis and Machine Intelligence},37(1):175-188(2015).
\bibitem{ssvm1}T. Joachims, Transductive inference for text classification using support vector machines. \emph{In Proc. 16th Int. Conf. On Mach. Learn., Bled, Slovenia,}  200-209(1999).
\bibitem{ssvm2} L.M. Yang, Q.Y. Su, B.Y. Yang, D. Tong, and X. Xiao,  A new semi-supervised support vector machine classifier based on wavelet transform and its application in the iris image recognition.,\emph{ Int J Appl Math Stat} 52(5):86-93(2014).
\bibitem{ssvm3} K. Lu, X. He, J. Zhao, Semi-supervised support vector learning for face recognition. \emph{Advances in neural networks-ISNN}, Springer, Berlin, 104-109(2006).
\bibitem{ssvm4} J. C. Ang, H. Haron, H. N. A. Hamed , Semi-supervised SVM-based feature selection for cancer classification using microarray gene expression data. \emph{In M. Ali., Y. Kwon, CH. Lee , J. Kim, K. Kim (eds) Current Approaches in Applied Artificial Intelligence IEA/AIE 2015, Seoul, South Korea,Lecture Notes in Artificial Intelligence-ISNN}, Springer, Cham., 9101,468-477(2015). 
\bibitem{svmp1}C. Cortes, V. Vapnik, Support-vector networks, \emph{Mach. Learn.}, 20(3), 273-297(1995).
\bibitem{svmp2}S. Abe, Support Vector Machines for Pattern Classification. Berlin, Germany: Springer-Verlag(2005).
\bibitem{s3vm} O. Chapelle, A. Zien, Semi-supervised learning by low density separation. In Proc. of the 10th Int. Wksh. on Arti. Intell. and Stat., Savannah,U.S.A., 57-64(2005).
\bibitem{Zhu} X. Zhu, Semi-Supervised Learning Literature Survey, \emph{technical report}, Univ. of Wisconsin-Madison, 2007
\end{thebibliography}
\end{document}